\documentclass[%
reprint,
superscriptaddress,
aps,
prb,
]{revtex4-2}

\usepackage{graphicx}% Include figure files
\usepackage{dcolumn}% Align table columns on decimal point
\usepackage{bm}% bold math
\usepackage{physics}
\usepackage{amssymb}
\usepackage[caption=false]{subfig}
\usepackage{color}
\usepackage{ulem}
\usepackage{comment}
\usepackage{enumitem}
\usepackage[squaren]{SIunits}
\usepackage{textgreek}
\usepackage{soul}

\usepackage{hyperref}
\hypersetup{colorlinks=true,linkcolor=blue,citecolor=blue,urlcolor=blue}

\newcommand{\D}{\mathrm{d}}

\definecolor{darkgray}{gray}{0.4}
\begin{document}

%\preprint{APS/123-QED}

\title{Localization structure of electronic states in the quantum Hall effect}

\author{Alioune Seye}
\affiliation{
Laboratoire de Physique de la Matière Condens\'ee, Ecole Polytechnique, CNRS, Institut Polytechnique de Paris, 91120 Palaiseau, France
}

\author{Marcel Filoche}
\affiliation{
Institut Langevin, ESPCI Paris, Universit\'e PSL, CNRS, 75005 Paris, France
}

\date{\today}

\begin{abstract}
We investigate the localization of electronic states in the integer quantum Hall effect using a magnetic localization landscape (MLL) approach. By studying a continuum Schrödinger model with disordered electrostatic potential, we demonstrate that the MLL, defined via a modified landscape function incorporating magnetic effects, captures key features of quantum state localization. The MLL effective potential reveals the spatial confinement regions and provides predictions of eigenstate energies, particularly in regimes where traditional semiclassical approximations break down. Numerical simulations show that below a critical energy, states localize around minima of the effective potential, while above it, they cluster around maxima—with edge effects becoming significant near boundaries. Bridging the gap between semiclassical intuition and full quantum models, the MLL offers a robust framework to understand transport and localization in disordered quantum Hall systems, and extends the applicability of landscape theory to magnetic systems.
\end{abstract}

\maketitle

%\tableofcontents

\section{Introduction}

The integer quantum Hall effect (IQHE) is characterized by the quantization of Hall conductivity, which takes precise values independent of the intensity of the magnetic field~\cite{Klitzing_1980_PhysRevLett}. To arise, this quantization requires a strong magnetic field to quantize the kinetic energy, forming Landau levels, and the presence of a quenched disordered potential inside the material, responsible for localizing almost all bulk eigenstates. The remaining conducting states are either bulk or edges states. In the bulk, delocalization occurs around critical energies---one per Landau level---corresponding to non vanishing longitudinal conductivity in the zero temperature limit. Above each critical energy, one family of edge states (one per Landau level) yields a conductivity quantum of the transverse conductivity, explaining its quantized value as a multiple of $e^2/h$.

To analyze the effect of the quenched disorder on the localized states, two main directions have been explored. The first one consisted of modeling the potential as a random Gaussian white noise in field theories, and deriving an effective theory in which the longitudinal and transverse conductivities appear as parameters~\cite{Pruisken_1984_NuclPhysB}. Quantization of the conductivity emerges as fixed points of the renormalization group flow, the critical energies being associated with transitions. The second direction consisted of studying the semiclassical limit in which the disorder is smooth at the magnetic length scale, allowing for a simpler description of the eigenstates and their localization properties. Electronic transport then occurs through a network formed by equipotentials of the disorder potential connecting its saddle points~\cite{Chalker_1988_JPhysC, Huckestein_1995_RevModPhys, Kramer_2005_PhysRep}.

However, neither approach can completely explain experimental results. The critical exponent describing the divergence of the localization length is the focus of intensive research, and a wide variety of values have been observed experimentally in different systems, as well as being predicted by various theoretical models. In 2005, Li \textit{et al.} showed that in heterostructures made of disordered alloys, different concentrations of the alloying metal give different exponents, with no clear universal behavior for all concentrations~\cite{Li_2005_PhysRevLett}, an observation consistent with earlier results~\cite{Diorio_1992_High,Gudina_2018_Semiconductors, Balaban_1998_PhysRevLett}. In 2009, Slevin and Ohtsuki proved that previously computed exponents in the widely used Chalker-Coddington network model did not account for finite-size scaling effects. Yet, their computed exponent was not in agreement with the experiments either~\cite{Slevin_2009_PhysRevB}. Gruzberg \textit{et al.} studied a geometrically disordered network model which gave an exponent closer to the experimental measurements~\cite{Gruzberg_2017_PhysRevB}. The authors also showed that other network models thought to be in the same universality class yield different exponents~\cite{Dresselhaus_2021_AnnPhys}, questioning even more the universality of the two-parameter scaling~\cite{Balaban_1998_PhysRevLett, Arapov_2019_LowTempPhys, Gudina_2018_Semiconductors, Dodoo-Amoo_2014_CondensMatterPhys}. Zirnbauer analyzed the renormalization group flow of a conformal field theory describing the transition in such network models, demonstrating that model-dependent finite-size (marginal or irrelevant) effects persist over large system sizes~\cite{Zirnbauer_2019_NuclPhysB}. In addition, most numerical studies either rely on transfer matrix methods which are not really 2D systems---as they are based on a dimensional crossover to one dimension---or assume a large magnetic field limit by projecting on the lowest Landau level. Although it is widely admitted that the disorder characteristics have a significant impact on the transition~\cite{Diorio_1992_High, Gudina_2018_Semiconductors, Balaban_1998_PhysRevLett, Li_2005_PhysRevLett}, a specific theoretical study on this question, away from the large magnetic field limit, is still missing.
	
In this paper, we analyze the localization properties of quantum states in an IQHE continuum model using a localization landscape (LL) approach. This approach was introduced in \cite{Filoche_2012_PNAS} for non-magnetic Schrödinger operators. Among its many features, it predicts the location of the eigenstates~\cite{Filoche_2012_PNAS}, their long-range spatial decay~\cite{Arnold_2019_CommPDE},  the densities of states~\cite{David_2021_AdvMath}, and the structure of transportation networks in variable-range hopping models~\cite{Thayil_2023_ApplPhysLett}. Recently, Poggi proposed an extension of the~LL to magnetic Schrödinger operators and proved decay bounds on eigenfunctions via this magnetic localization landscape (MLL)~\cite{Poggi_2024_AdvMath}. Here, we show how this MLL helps to understand and predict the locations, the spatial structures, and the energies of the electronic eigenstates in an IQHE system. In the first section we describe our model. In the second section we present the MLL and study some specifics of this tool in our setting. In the third section we present our numerical analysis of the eigenstates compared to MLL predictions. Finally we discuss the benefits and limits of this approach.

\section{The physical model}

\subsection{The semiclassical picture}

Let us review the main arguments of the semiclassical approach to localization in the IQHE. Classically, an electron subject to a transverse uniform magnetic field $\vb{B} = B\,\vb{e}_z$ and whose motion is constrained to a plane follows a cyclotron orbit. Its energy $E$ is entirely kinetic:
\begin{equation}\label{eq:cyclotron_energy}
E=\frac{1}{2} m_e R^2 \, \omega_c^2 \,,
\end{equation}
where $m_e$ is the mass of the electron, $R$ the radius of the orbit, and $\omega_c=e\abs{B}/m$ is the cyclotron angular frequency. Adding a uniform electric field~$\vb{E}$ modifies this circular orbit by composing it with a drift at uniform velocity $\vb{v_D}= \vb{E} \cross \vb{B}/B^2$. If the electric field is not uniform but derives from an electrostatic potential~$V$ whose spatial variations occur at a length scale much larger than the size of the orbit, it can be seen as locally uniform at this scale: in this case the electron orbit drifts with the local velocity $\vb{v_D}$.

In the quantum formalism and in the absence of an electric field, the electronic energies are quantized and are akin to the ones of a quantum harmonic oscillator, i.e., regularly spaced:
\begin{align}\label{Landau_energies}
E_{n,B}= \left(n+\frac{1}{2}\right)\hbar\omega_c =\frac{1}{2} m_e (2n+1) \, \ell_B^2 \, \omega_c^2 \,,
\end{align}
where $\ell_B= \displaystyle \sqrt{\hbar/(e\abs{B})}$ denotes the magnetic length. Each of these so-called Landau levels is highly degenerate. Due to degeneracy, the eigenfunctions of a Landau level can take a wide variety of shapes. Simply adding a uniform electric field lifts the degeneracy in all Landau levels. In that case, the eigenstates of the nth Landau level are confined along infinite strips in the direction of $\vb{v_D}$, and their wave functions are products of a plane wave of wave number $k$ in the direction of the strip and the nth eigenstate of a harmonic oscillator in the transverse direction, centered on an abscissa which depends on $k$.

The eigenstates are thus localized around the electrostatic equipotential lines. Their energy is given by the quantized Landau energy $E_{n,B}$ plus the potential energy at the center of the strip. In the semiclassical limit, i.e., in the presence of an electrostatic field $\vb{E}$ varying on length scales much larger than $\ell_B$, the electric field is locally almost constant at the magnetic scale. In this regime, the eigenstates exhibit the aforementioned features:
\begin{enumerate}[label=(\arabic*)]
\item They form strips localized around equipotential lines and have energies given by the Landau level energy plus the potential at the center of the strip.
\item In a disordered system with a smooth potential varying on length scales large compared to~$\ell_B$, almost all equipotential lines are closed. Consequently, most states are effectively localized and do not contribute to the conductivity.
\item There is one single energy at which the equipotential lines can percolate through the sample. Delocalized eigenstates appear in each Landau level at specific energies corresponding to this percolation energy plus $E_{n,B}$. 
\end{enumerate}

This semiclassical picture has been widely used to predict the exponents of the transition between conductivity plateaus. The network model~\cite{Chalker_1988_JPhysC} uses the semiclassical picture to associate a complex scattering amplitude to the junctions of equipotentials. It predicts the divergence of the eigenstates localization length at a critical energy, and a non-vanishing longitudinal conductivity as the Fermi level reaches this critical energy. 

\subsection{The intermediate regime}

The disordered single-particle quantum Hall system is characterized by three competing length scales, each length scale $\ell$ being associated to an energy through the relation $E=\hbar^2/(2m_e \ell^2)$:
\begin{enumerate}[label=(\arabic*)]

\item The first length scale is the aforementioned magnetic length $\ell_B=\sqrt{\hbar/(e\abs{B})}$: it provides the typical width of the semiclassical strip-like eigenstates. The associated energy is the energy of the first Landau level $E_B=\hbar e\abs{B}/(2m_e)$.

\item The second length scale~$\ell_V$ is determined by the intensity of the disorder. It can be seen as the typical (or smallest) localization length for the system in the absence of magnetic field. For bounded disordered potentials, this disorder length is $\hbar/ \sqrt{2m_e\overline{V}}$ where $\overline{V}$ is the typical fluctuation of the potential. In our study, it is computed as the potential average minus the minimum of V. The associated energy is simply~$\overline{V}$.

\item The third relevant length scale~$\ell_{\textrm{corr}}$ is the correlation length scale of the disordered potential. The associated correlation energy is $E_{\textrm{corr}}=\hbar^2/(2m_e \ell_{\rm corr}^2)$.

\end{enumerate}

\begin{figure}
	\centering
	\includegraphics[width=0.8\columnwidth]{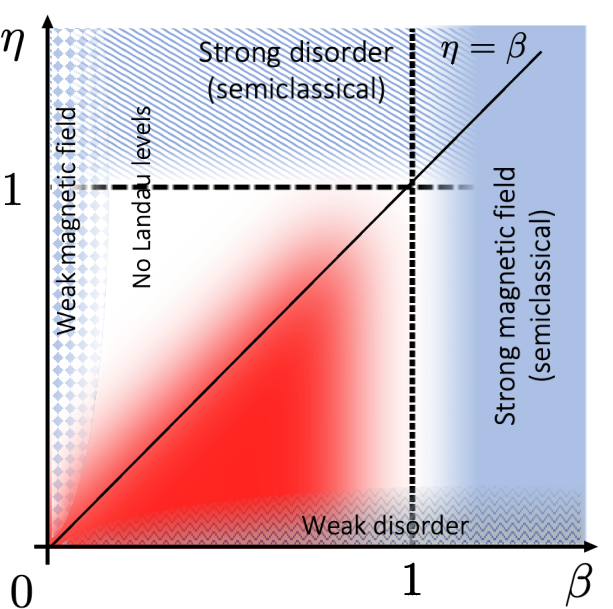}
	\caption{Different regimes of disorder and magnetic intensity in the parameter space $(\beta,\eta)\approx (\ell_{\rm corr}^2/\ell_B^2,\ell_{\rm corr}^2/\ell_V^2)$. The regime of interest in this work is symbolized by the red hue.}\label{fig:regimes}
\end{figure}

Depending on the relative values of these three scales, several main regimes emerge, summarized in Fig.~\ref{fig:regimes}:
\begin{enumerate}[label=(\arabic*)]

\item A semiclassical regime of confinement by the potential when $\ell_V \ll \ell_{\rm corr}$ and $\ell_V \ll \ell_B$. 

\item A semiclassical regime of strong magnetic field when the correlation length is larger than the magnetic length, i.e., $\ell_B \ll \ell_{\rm corr}$ as discussed above.

\item When $\ell_{\rm corr} < \ell_V \ll \ell_B$, the magnetic field is weak and the correlation length is the smallest length scale. This corresponds to a slight magnetic perturbation of the strong Anderson-localized regime. In a 2D disordered non-magnetic system, all states are localized with a localization length which grows exponentially with the energy. Hence, only the states initially localized over regions that are of order of the magnetic length or larger are substantially modified by the magnetic field. Delocalized states appear at high energies~\cite{Khmelnitskii_1984_PhysLettA, Laughlin_1984_PhysRevLett, Yang_1996_PhysRevLett}.

\item An intermediate regime investigated in this work in which the system size $L$ is significantly larger than all three characteristic lengths, and
\begin{equation}\label{eq:intermediate_regime}
	\ell_{\rm corr} \lesssim \ell_B \lesssim \ell_V \, .
\end{equation}

\end{enumerate}
In this regime, Landau levels remain separate since $\ell_{B} \lesssim \ell_V$, and the semiclassical picture is not relevant since $\ell_{\rm corr} \lesssim \ell_B$.  The localization properties are largely affected by the magnetic field. This magnetic field-driven localization regime occurs in quantum Hall devices where disorder fluctuations occur on a few atomic scales~\cite{Chalker_1999_AspectsTopologiques}, and for magnetic fields of several tesla such that the magnetic length~$\ell_B$ is below \unit{10}{\nano \meter}. A description of localization properties of eigenmodes is still lacking in this setting which is addressed via our approach.

\subsection{The mathematical model}

A 2D gas of non-interacting charged particles in a uniform magnetic field $\vb{B}=B\ \vb{e}_z$ and a (disordered) potential $V$ is described by the following single-particle Schrödinger equation:
\begin{equation}
	i\hbar \,\partial_t\psi = \hat{H} \psi = \frac{1}{2m_e}\left(-i\hbar\grad +e\vb{A}\right)^2\psi+V \psi \,,
\end{equation}
where $m_e$ is the electron effective mass, $-e$ its charge, and $\vb{A}$ is a magnetic vector potential satisfying $\curl\vb{A} = \vb{B}$. We can choose $B>0$ without loss of generality.

We study a type of potential that represents energy band fluctuations as a result of compositional alloying. This choice is inspired by quantum Hall experiments performed on AlGaAs semiconductor interfaces~\cite{Li_2005_PhysRevLett}. To generate the potential, we first assign a Boolean variable (0 or 1, mimicking the presence of an Al or Ga atom) on a grid of lattice parameter~$a$. We then compute a~local effective alloy content by convoluting this Boolean variable with a Gaussian function of standard deviation $\sigma \approx 2a$.

The eigenvalue equation is rescaled by considering a dimensionless variable $\vb{r}/2a$:
\begin{equation}\label{eq:dimensionless}
	\left(-i\grad+ \frac{E_B}{E_{2a}}\vb{A} \right)^2\psi+\frac{V}{E_{2a}}\psi=\frac{E}{E_{2a}}\psi\,,
\end{equation}
where $E_{2a}= \hbar^2/(2m_e(2a)^2)$, $E_B=\hbar eB/2m_e$  is the first Landau energy in Eq.~\eqref{Landau_energies} and $\grad \times \vb{A} = \, \vb{e}_z$. The problem is then parameterized by dimensionless quantities: the magnetic intensity $\beta=E_B/E_{2a}$, the disorder strength $\eta=\overline{V}/E_{2a}$ and the non dimensional energy $E/E_{2a}$. We solve the following eigenvalue problem:
\begin{equation}\label{eq:nondimensional_eigvv}
	\left(-i\grad + \beta \vb{A}\right)^2\psi +\eta V\psi = E\psi \,,
\end{equation}
in a box of size $L/2a=200$ with Dirichlet boundary conditions. The disordered non dimensional potential $V$ has mean value~1. Our analysis is based on simulations in which the parameters take the following values: $\eta=0.001,\, 0.2$, and $\beta=0.05,\, 0.1,\, 0.2$. The corresponding dimensionless magnetic and disorder lengths are $\ell_V\approx 31.6,\, 2.2$, and $\ell_B \approx 4.5,\, 3.2,\, 2.2$, respectively. We thus investigate a strong disorder and a weak disorder regime for different values of an intermediate magnetic intensity. 

\section{Numerical methods}

\subsection{Gauge-invariant finite element method}

In Maxwell's equations, electric and magnetic fields can be defined up to a gauge transformation. Such a gauge transformation affects only the phase of the electronic wavefunctions. To show it, consider a non dimensional potential vector~$\vb{A}$ such that $\curl{\vb{A}}=\vb{e}_z$, and $\hat{H}=\left(-i\grad + \beta \vb{A}\right)^2 +\eta V$ the associated Hamiltonian. If $\psi$ is an eigenfunction of~$\hat{H}$ with energy $E$, then $\tilde{\psi}=\exp(i\phi) \,\psi$ is an eigenfunction of the same energy~$E$ of the Hamiltonian $\hat{H}_\phi = (-i\grad + \beta \tilde{\vb{A}} )^2 +\eta V$, with $\beta\tilde{\vb{A}} = \beta\vb{A} - \grad \phi$ (hence of identical magnetic field), $\phi$ being any smooth function:
\begin{align}
	\hat{H}_\phi \tilde{\psi} & = \left[(-i\grad + (\beta\vb{A} -\grad \phi))^2+\eta V\right]\exp(i\phi)\ \psi \nonumber\\
		&= \exp(i\phi) \left[(-i\grad + \beta\vb{A})^2+\eta V\right]\psi \nonumber\\
		&= \exp(i\phi) \hat{H}\psi = E\, \tilde{\psi}.
\end{align}
Consequently, there is a one-to-one correspondence between all eigenstates of two different gauges, up to a local phase factor.

This gauge dependency plays an important role when it comes to simulating the eigenvalue problem. In particular, high-frequency fluctuations of the phase of the eigenstates related to the choice of gauge make non gauge-invariant methods difficult to exploit. High-order finite elements have shown to be more accurate to capture the phase fluctuations of the eigenstates~\cite{Bonnaillie-Noel_2007_CMAME}. However, we are not interested in the non-physical phase fluctuations, but rather in quantities involving $(-i\grad +\beta \vb{A})\psi$. Therefore, we implemented a gauge-invariant finite element method on a triangular mesh~\cite{Christiansen_2011_SIAMJNumerAnal}. The numerical scheme is based on a discretization of $\left(-i\grad+\beta\vb{A}\right)^2$ using a phase factor to account for the magnetic part similar to the Peierls substitution in a lattice:
\begin{align}
    &\mel{\varphi}{(-i\grad+\beta\vb A)^2}{\psi} \nonumber\\
    &\quad \approx \sum_{i\sim j}\ \mu_{ij}\left(\varphi_j-\mathrm e^{i \beta A_{ij}}\varphi_i\right)^*\left(\psi_j-\mathrm e^{i \beta A_{ij}}\psi_i\right) \,,
\end{align}
with $A_{ij}=\int_{i\rightarrow j}\vb A\cdot \D \vb l$ and $\mu_{ij}$ a geometrical factor, the sum being performed on all $(i, j)$ neighboring vertices of the mesh. This method was shown to exhibit good convergence properties and to be insensitive to phase fluctuations possibly induced by the gauge. In particular, for high magnetic intensities, phase fluctuations at the scale of the mesh size do not necessarily affect the performance of the method, although some of the phase fluctuations may not be visible for a particular gauge choice. This method is sufficient for one particular gauge which we choose to be the symmetric gauge: $\vb{A}= -y/2\, \vb{e}_x + x/2 \, \vb{e}_y$.

\subsection{The underestimation of the kinetic energy}
	
In the weak disorder regime ($\eta \ll 1$), the energy of an eigenstate is mostly kinetic. Computing the energy therefore requires us to assess precisely this kinetic contribution. The spatial oscillations of the eigenstate wave function occur at a typical length scale that is the magnetic length scale~$\ell_B$. Consequently, for a uniform mesh refined at a scale smaller than $\ell_B$, and thanks to the gauge invariance of our numerical method, the estimation of the kinetic energy is independent on the specific location of the eigenstate in the domain. Therefore, the discrepancy between the actual energy of a state and the computed one only depends on the mesh discretization and the magnetic intensity, and is almost identical among all states. We evaluated this discrepancy~$\delta E$ between the minimum kinetic energy $E_{\rm k, min}$ and the value~$\beta$ (the theoretical value of the minimum) on the first 100~eigenstates of the system for five disorder realizations for $\eta=10^{-3}$:
\begin{enumerate}[label=(\arabic*)]
	\item  \makebox[2cm][l]{$\beta=0.2$:} $\delta E = E_{\rm k, min} - \beta \approx -2.1(5)\times 10^{-4}$,
	\item  \makebox[2cm][l]{$\beta=0.1$:} $\delta E \approx -5.3(5)\times 10^{-5}$,
	\item  \makebox[2cm][l]{$\beta=0.05$:} $\delta E \approx -1.3(0)\times 10^{-5}$.
\end{enumerate}
The last digit was common to all disorder realizations. The energy spreading of the Landau levels being about $4\times10^{-4}$ for all three values of $\beta$, the underestimation~$\delta E$ due to the discretization has to be accounted for. In the following, unless otherwise specified, we compensated this systematic shift by adding~$\delta E$ to all energies.

\subsection{Assessing the localization length}

\begin{figure}[ht!]
	\centering
	\includegraphics[width=\columnwidth]{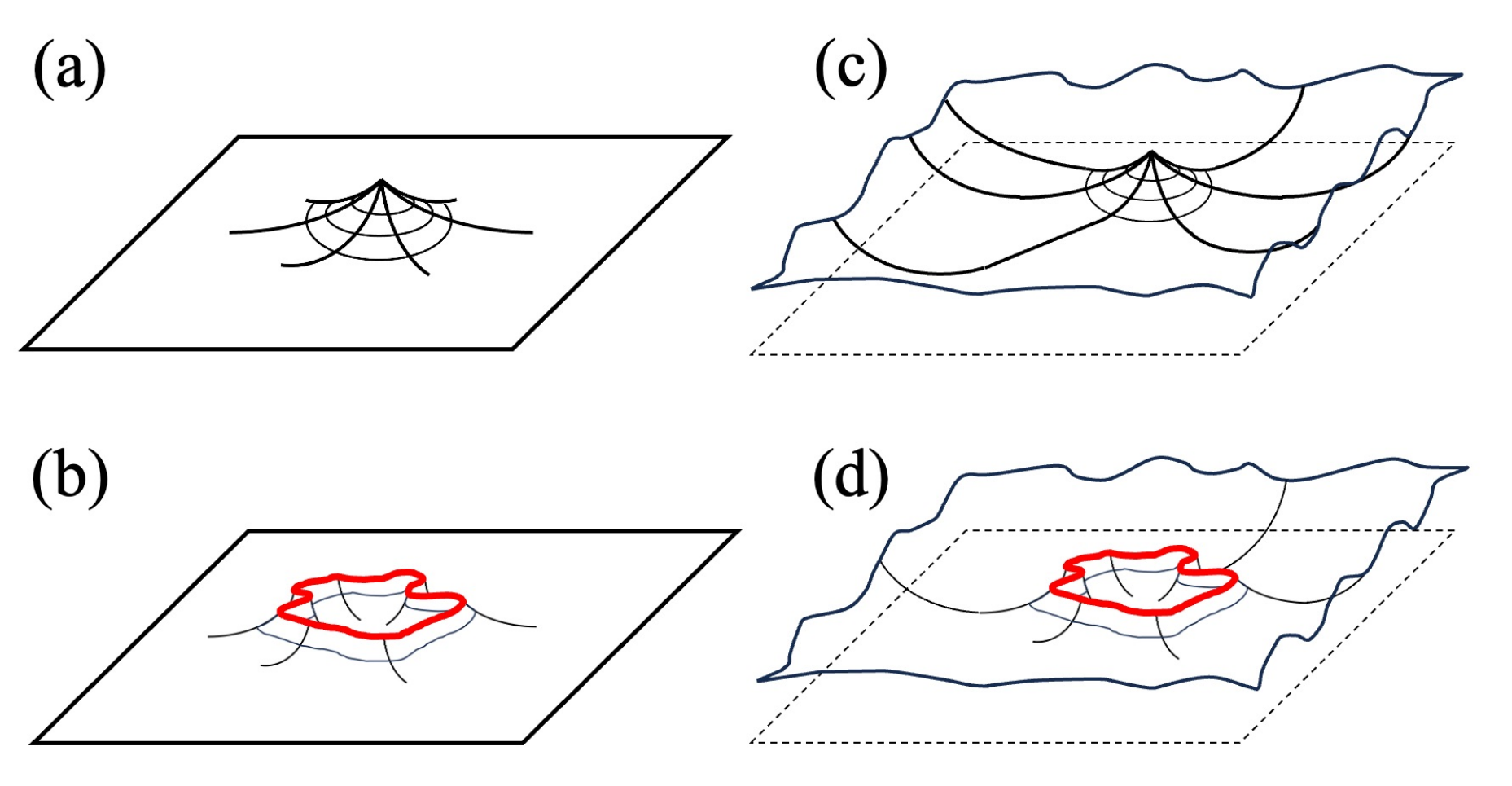}
	\caption{Typical shapes of $\abs{\psi}^2$ for localized eigenfunctions in the IQHE: (a) and (b) are wave functions localized inside the bulk, both below the critical energy, with no contribution at the edges of the domain. Type (a) correspond to a local fundamental state at the bottom of a potential well, whereas the red line in (b) is a semiclassical trajectory (a level set of the effective potential) for a higher-energy state. (c) and (d) are states above the critical energy. They look similar to (a) and (b), except for a significant contribution at the edges.}
	\label{fig:eigenfunction_types}
\end{figure}

One needs to design a method able to compute a localization length that characterizes the spatial decay of any exponentially localized state, even in the absence of rotational symmetry. To that end, for any wave function~$\psi(\vb{r})$ we define the function~$\mathcal{A}_\psi(\varepsilon)$ which measures the area where the probability density~$\abs{\psi}^2$ is larger than~$\varepsilon$:
\begin{equation}
\mathcal{A}_\psi(\varepsilon) = \int_{\mathbb{R}^d}\Theta\left(\abs{\psi(\vb{r})}^2-\epsilon\right) ~\D^d\vb{r} \,,
\label{eq:A_psi}
\end{equation}
$\Theta$ being the Heaviside function. We then assess the behavior of that area function for two types of localized eigenstates: first, a localized state inside the bulk, second, a state localized near the edges of the domain (see Fig.~\ref{fig:eigenfunction_types}). We assume in both cases that the typical decay length is much smaller than the size of the domain.

First, if the probability density decays exponentially with a characteristic length~$\ell$ away from a central location~$\vb{r}_0$ inside the bulk:
\begin{equation}\label{eq:ansatz_bulk}
\abs{\psi(\vb{r})}^2 = C \, \exp(-\frac{\abs{\vb{r}-\vb{r}_0}}{\ell}) \,,
\end{equation}
then $\mathcal{A}_\psi(\varepsilon)$ can be exactly computed:
\begin{align}
\mathcal{A}_\psi(\varepsilon) &= \int\limits_{A \, e^{-\frac{\abs{\vb{r}-\vb{r}_0}}{\ell}} > \varepsilon} \D^d\vb{r} = \, \omega_d \, \left( \ell \ln\left( \frac{C}{\varepsilon} \right)\right)^d \,,
\end{align}
where $\omega_d$ is the volume of the unit ball in dimension~$d$. In other words:
\begin{equation}\label{eq:bulk_fit}
\left[\frac{\mathcal{A}_\psi(\varepsilon)}{\omega_d}\right]^{1/d} = -\ell \ln(\varepsilon) + \ell \ln(C) \,.
\end{equation}
In this case we compute a localization length~$\ell$ as the slope of the linear regression of $\left[\mathcal{A}_\psi(\varepsilon)\right/\omega_d]^{1/d}$ vs. $-\ln(\varepsilon)$ (in our study, $d=2$ and $\omega_2 = \pi$). This corresponds to Figs.~\ref{fig:eigenfunction_types}(a) and \ref{fig:eigenfunction_types}(b).

Second, if the probability density decays exponentially away from the boundary~$\Gamma$ of the domain with a a characteristic length~$\ell$, then:
\begin{equation}\label{eq:ansatz_edge}
\abs{\psi(\vb{r})}^2 = C \, \exp(-\frac{d(\vb{r},\Gamma)}{\ell})\,,
\end{equation}
where $d(\vb{r},\Gamma)$ is the distance to the boundary~$\Gamma$. In $d=2$ the area function takes the form:
\begin{align}
    \mathcal{A}_\psi(\varepsilon) &= \int\limits_{C \, \exp(-\frac{d(\vb{r},\Gamma)}{\ell}) > \varepsilon} \D^d\vb{r} \nonumber\\
    &= L^2-\left(L-2\,d(\varepsilon)\right)^2\,,
\end{align}
where $d(\varepsilon)$ is the distance to the boundary where the probability density has value $\varepsilon$. In other words:
\begin{equation}\label{eq:edge_fit}
d(\varepsilon) = \frac{L-\sqrt{L^2-\mathcal{A}_\psi(\varepsilon)}}{2} = -\ell \ln(\varepsilon) + \ell \ln(C) \,.
\end{equation}
We compute a localization length~$\ell$ as the slope of the linear regression of $\left(L-\sqrt{L^2-\mathcal{A}_\psi(\varepsilon)}\right)/2$ vs. $-\ln(\varepsilon)$. This corresponds to Figs.~\ref{fig:eigenfunction_types}(c) and \ref{fig:eigenfunction_types}(d).

In all cases, we favored this area function over the widely-used inverse participation ratio (IPR) because the~IPR essentially measures the size of the main exis-tence region of the state, whereas we are interested in determining the long-range decay of the quantum state (the IPR and the area method providing the same estimate in the first case of an exponentially decaying state in the bulk).

\begin{figure*}[ht!]
	\centering
	\includegraphics[width=.9\textwidth]{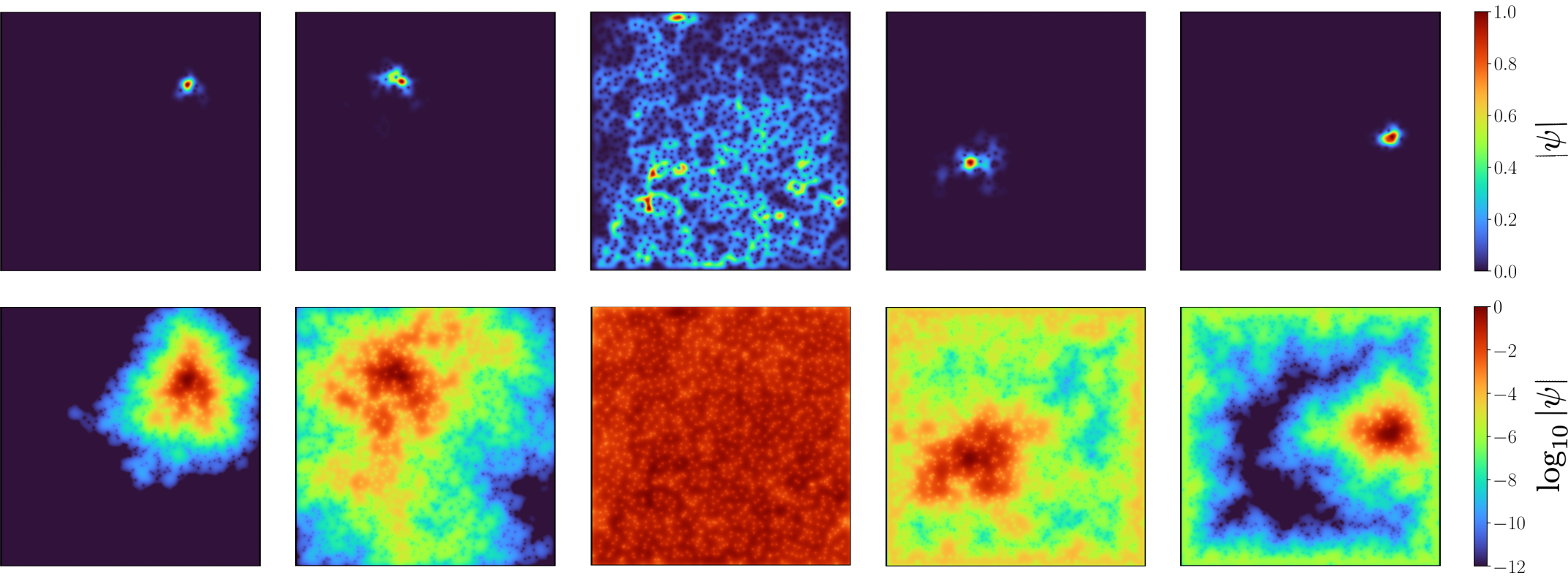}\\
	\raggedright
	\hskip 16mm $E = 0.334$ \hskip 14.5mm $E=0.361$ \hskip 14.5mm $E=0.397$ \hskip 14.5mm $E=0.442$ \hskip 14.5mm $E=0.462$ 
	\caption{Color maps of the moduli of five eigenstates of Eq.~\eqref{eq:nondimensional_eigvv} with parameters $(\beta, \eta) = (0.2 , 0.2)$, inside the first Landau level (top row, linear scale, bottom row, logarithmic scale). From left to right, the energies are about $0.334,\, 0.361,\, 0.397,\, 0.442,\, 0.462$, respectively. One observes that the first two states are localized, similarly to Figs.~\ref{fig:eigenfunction_types}(a) and \ref{fig:eigenfunction_types}(b), respectively, the center state is clearly delocalized, and the last two states are localized again, albeit with a significant probability density on the edges of the domain, similar to Figs.~\ref{fig:eigenfunction_types}(d) and \ref{fig:eigenfunction_types}(c), respectively. All localized states have a localization length of the order of few units, making the decay of the wave function amplitude barely visible on a linear scale.}
	\label{fig:eigenstates_lin-log}
\end{figure*}

\section{The spatial and spectral structure of magnetic states}

\subsection{The localization properties}

Using our gauge-invariant finite element method, we compute the eigenstates in the first broadened Landau level. In Fig.~\ref{fig:eigenstates_lin-log} we display the moduli of five eigenstates, for large magnetic field---within the studied range---and large disorder, $(\beta, \eta)=(0.2, 0.2)$, in linear (top row) and logarithmic (bottom row) scale. Energies range within the first Landau level (from $E\approx 0.34$ to $E\approx 0.46$). The first two states are well localized and exhibit a clear decay from the principal peak of the wave function, whereas the third one spans across the whole system, showing delocalization. The two states at higher energy not only exhibit an exponential decay away from their main localization region in the bulk, but also have a fraction of their probability density located on the edges of the domain (also decaying exponentially away from the edges).

\begin{figure}[h!]
	\centering
	\includegraphics[width=0.9\columnwidth]{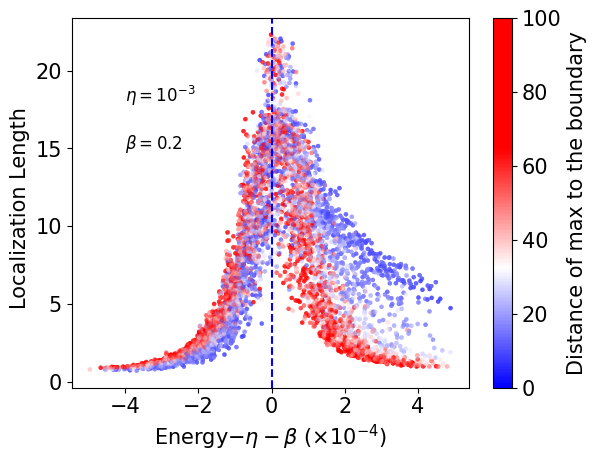}\\
	\includegraphics[width=0.9\columnwidth]{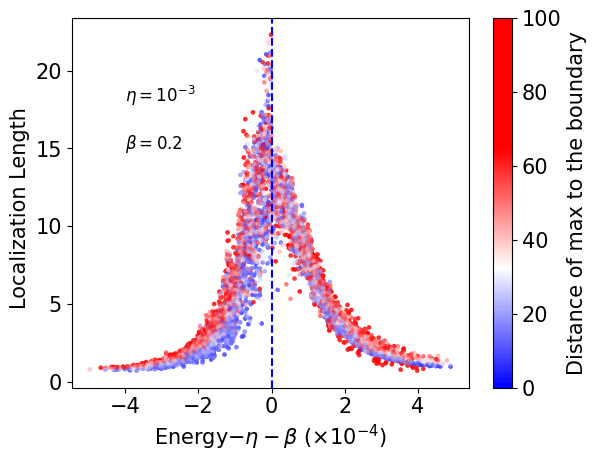}
	\caption{Top: Localization lengths of the first Landau level as a function of energy for $\eta=0.001$, $\beta=0.2$, for five potential realizations. The system size is $L=200$. The blue to red hue corresponds to the distance of the peak of the wave function to the boundary. The vertical dashed line materializes the energy $E_c=\eta+\beta$ which appears to be the critical energy. The localization length is assessed from the behavior of~$\mathcal{A}_\psi(\varepsilon)$, assuming an exponential decay from the peak of the wave function, see Eqs.~\eqref{eq:A_psi} and \eqref{eq:bulk_fit}. Bottom: Similar plot as above, except that the localization length is now extracted from Eq.~\eqref{eq:edge_fit} for all eigenstates of energy larger than~$E_c$, theses states having a significant probability density at the edges of the domain.}
	\label{fig:loc_lengths}
\end{figure}

In Fig.~\ref{fig:loc_lengths} (top), we show the computed localization lengths (see Eq.~\eqref{eq:ansatz_bulk} of the first Landau level for five different realizations of the random potential, for the parameters values~$(\eta, \beta)=(0.001, 0.2)$. For each state, the color of the dot indicates the distance of its maximum probability density~$\vb{r}_{\rm max}$ to the boundary. On the lower part of the Landau level, the localization length flattens around its ground-state value. The localization length exhibits a non monotonous behavior as the energy increases: it first increases and then decreases throughout the Landau level, reaching a large fraction of the system linear size in the middle of the Landau level. Actually, the localization length is expected to diverge at a given critical energy in the Landau level~\cite{Levine_1984_NuPhysB}, which is located theoretically very close to~$E_c=\beta + \eta$. It should be emphasized that too much importance should not be attached to the exact value of the localization length near the critical energy (this value being here around 20) because formulas \eqref{eq:bulk_fit} and \eqref{eq:edge_fit} lose their accuracy close to the critical energy as delocalized states do not satisfy neither ansatz~\eqref{eq:ansatz_bulk} nor~\eqref{eq:ansatz_edge}.

The method for extracting the localization length assumes an approximate exponential decay of the wave function from a peak located in the bulk of the system; see Eq.~\eqref{eq:bulk_fit}. Our computations show that the values found are consistent for states inside the bulk (in red) and for states located near the edges of the domain (in blue) below the critical energy. However, this ansatz loses its validity above the critical energy, ${\sim}\beta+\eta$, because a significant fraction of the eigenstate probability density is to be found near the edges of the domain. This is the origin of the splitting of the right part of the graph in two branches, in the top frame of Fig.~\ref{fig:loc_lengths}. We thus introduced the ansatz~\eqref{eq:ansatz_edge} for eigenstates above the critical energy, leading to a slightly different method for extracting the localization length for these states, see Eq.~\eqref{eq:edge_fit}. Figure~\ref{fig:loc_lengths} (bottom) shows that this approach offers consistent results below and above the critical level, both for bulk and edge states.

\subsection{The localization landscape approach}

\subsubsection{Properties of the non-magnetic localization landscape}
\label{subsubsec:non-magnetic landscape}

The LL is a theoretical tool introduced in 2012 to investigate the localization properties of vibrating systems, especially quantum systems~\cite{Filoche_2012_PNAS}. Let us consider the non dimensional eigenvalue problem of Eq.~\eqref{eq:nondimensional_eigvv} with vanishing magnetic field:
\begin{equation}\label{eq:eigenvv}
	\hat{H} \psi =-\Delta \psi + \eta V \psi= E \, \psi\,,
\end{equation}
where $\eta V$ is a non-negative potential. The LL $u$ is there defined as the solution to the linear problem 
\begin{equation}
	\hat{H} u=1\,.
\end{equation}
The first property of this LL is that it provides a point-wise linear control on the modulus of any eigenfunction~$\psi$ of energy~$E$~\cite{Filoche_2012_PNAS}:
\begin{equation}\label{eq:linear_control}
	\abs{\psi (x)} \leq E \, \norm {\psi}_\infty u(x)\,.
\end{equation}
Wherever $u(x)\leq 1/E$, this bound applies a constraint on the modulus of~$\psi$. In particular at low energies, the localized eigenstates are essentially supported in a finite region which corresponds to one of the hills of the LL.

One of the major features of the LL is that its reciprocal, $1/u$, plays the role of an effective potential~\cite{Arnold_2016_PhysRevLett}. This become clear when conjugating the Hamiltonian with~$u$ (i.e., looking at the properties of $u^{-1} \hat{H} u$). This operation brings a different decomposition of the average energy of any quantum state~$\psi$ as the sum an effective kinetic energy and an effective potential energy:
\begin{equation}\label{eq:conjugate_energy}
		\int \abs{\grad{\psi}}^2+\eta V\abs{\psi}^2 = \int u^2 \,\abs{\grad \left(\frac{\psi}{u}\right)}^2 + \int \frac{1}{u}\abs{\psi}^2 \,.
\end{equation}

This equality also implies an exponential control on the spatial decay of any eigenfunction of energy~$E$ localized around~$\vb{r}_0$, via the so-called Agmon distance~$\rho$ computed using the effective potential~\cite{Arnold_2016_PhysRevLett, Arnold_2019_CommPDE}:
\begin{align}\label{eq:Agmon_non_mag}
&\abs{\psi(\vb{r})} \lesssim e^{-\rho(\vb{r}_0, \vb{r}; E)} \nonumber \\
&\textrm{with} \quad \rho(\vb{r},\vb{r'};E) = \min_{\gamma} \left[ \int\limits_{\gamma} \sqrt{\left(\frac{1}{u}-E\right)_+} ~{\rm d}s \right]\,,
\end{align}
the Agmon distance $\rho(\vb{r},\vb{r'};E)$ being defined as the minimum over all possible paths connecting $\vb{r}$ to $\vb{r'}$ of the local metric~$\sqrt{(1/u-E)_+}$, the subscript $+$ denoting the positive part. It is important to note that at energies larger than the typical values of the effective potential, the region defined by $(1/u<E)$ percolates at long distance and the Agmon distance can remain small even for arbitrarily distant points. In this case, the control by the Agmon distance does not impose any exponential decay of the eigenfunction amplitude~\cite{Filoche_2024_PhysRevB}.

In the next subsection we introduce the MLL. We first discuss why a specific tool is needed for magnetic systems, then define the MLL and finally list some fundamental properties of this MLL in our setting.

\subsubsection{The magnetic localization landscape}

The existence of complex terms in the magnetic Hamiltonian of Eq.~\eqref{eq:nondimensional_eigvv} requires to rethink the concept of the LL if one wants to generalize it to magnetic systems. For any eigenstate~$\psi$ of this magnetic Hamiltonian, one can locally rewrite the eigenvalue equation by separating the modulus and phase of the eigenvector $\psi=\abs{\psi}e^{i\phi}$, $\abs{\psi}$ and $\phi$ being two differentiable fields. The Schr\"odinger equation
\begin{align}
\left(-i\grad + \beta \vb{A}\right)^2\abs{\psi}e^{i\phi} + \eta V\abs{\psi}e^{i\phi} = E  \abs{\psi}e^{i\phi} \,,
\end{align}
can then be decomposed into two real equations:
\begin{align}
	\begin{cases}
	 \abs{\psi} \, \div{(\beta \vb{A}+\grad{\phi})}+2(\beta\vb{A}+\grad{\phi})\vdot \grad{\abs{\psi}} = 0 \\
	 -\Delta \abs{\psi}+\left((\beta \vb{A}+\grad{\phi})^2 + \eta V \right) \abs{\psi} =  E \, \abs{\psi}\,.
	\end{cases}
\end{align}
We see from the second equation that the modulus $\abs{\psi}$ satisfies a non-magnetic Schr\"odinger eigenvalue equation with a modified potential~$(\beta \vb{A}+\grad{\phi})^2+\eta V$ which is larger than the original potential. Following the results of Sec.~\ref{subsubsec:non-magnetic landscape}, one possible choice of the LL for $\abs{\psi}$ would thus be the solution $u_1$ to
\begin{equation}
	-\Delta u_1 +((\beta \vb{A}+\grad{\phi})^2+\eta V) \, u_1 = 1\,.
\end{equation}
By maximum principle, $u_1$ is smaller than the non-magnetic LL~$u_0$, solution to
\begin{equation}\label{eq:adim_LL}
-\Delta u_0 + \eta Vu_0 =1 \,.
\end{equation}
Therefore, for any eigenfunction~$\psi$ of the magnetic problem, the control inequality~\eqref{eq:linear_control} still holds in which~$u_0$ is the non-magnetic~LL. However, the information provided by this LL is really poor in the relevant regime. We have seen that Eq.~\eqref{eq:linear_control} is meaningful only for $E$ smaller than~$1/u_0$. In the bulk, $1/u_0$ is always of order $\eta$ while, in the magnetic case, all energies are larger than~$\beta$. When the magnetic field is dominant, i.e., $\beta>\eta$, the bounds are therefore meaningless except close to the edges.

Recently, Poggi proposed a generalization of the LL for magnetic Schr\"odinger operators, here called MLL~\cite{Poggi_2024_AdvMath}. This MLL is defined via a non-magnetic landscape equation, but  takes into account the local magnetic field in the form of an additional potential. In our setting, the magnetic field is constant, so the MLL~$u$ is defined as the solution to:
\begin{equation}\label{eq:MLL}
	-\Delta u + \left(\eta V+\beta\right) u = 1 \,,
\end{equation}
with Dirichlet boundary conditions. Poggi also derived an energy inequality for the MLL:
\begin{align}\label{eq:energy_inequality}
	\int u^2 \,&\abs{\grad \left(\frac{\abs{\psi}}{u}\right)}^2 + \int \frac{1}{u}\abs{\psi}^2 \nonumber\\
	&\leq \int 2\abs{(-i\grad +\beta \vb{A})\psi}^2+\eta V\abs{\psi}^2 \,.
\end{align}
We want to stress an important point here. The non-magnetic case allows for an exact rewriting of the energy, see Eq.~\eqref{eq:conjugate_energy}. In the magnetic case, Poggi's approach leads only to a control inequality~\eqref{eq:energy_inequality} on the total energy, the difference with the non-magnetic situation being the factor of~2 appearing in front of the kinetic part on the right-hand side. Yet, this inequality allows Poggi to deduce an exponential control on the wave function amplitude, using an Agmon distance similar to the one defined in~Eq.~\eqref{eq:Agmon_non_mag} for the non-magnetic case.

One can assess the behavior of the MLL at large magnetic field, i.e., at large $\beta$: Assuming that the potential~$V$ is smooth enough, it is possible to expand the MLL in powers of~$1/\beta$ from Eq.~\eqref{eq:MLL}:
\begin{equation}
	u = \frac{1}{\beta}\left(1- \frac{\eta V}{\beta} + \frac{\eta^2V^2}{\beta^2} - \frac{\eta \Delta V}{\beta^2} + O\left(\beta^{-3}\right)\right),
\end{equation}
which gives the effective potential expansion:
\begin{align}\label{eq:semiclassical_expansion}
	\frac{1}{u} &= \beta \Big[1+\frac{\eta V}{\beta} + \frac{\eta^2V^2}{\beta^2} 
	-\frac{(\eta^2V^2- \eta \Delta V)}{\beta^2} + O\left(\beta^{-3}\right) \Big] \nonumber\\
	&= \beta+\eta V+ \frac{\eta \Delta V}{\beta} + O\left(\beta^{-2}\right) \,.
\end{align}

$\beta$ being a constant here, the spatial variations of $1/u$ are dominated by the variations of~$V$ at large $\beta$: the basins and hills of~$1/u$ coincide in space and depth with those of~$V$. Note that the shift~$\beta$ introduced in Eq.~\eqref{eq:MLL} is simply the energy of the first Landau level which is entirely of kinetic nature. Since in the strong magnetic field regime, the states are confined on equipotentials, their energy is simply the energy of their Landau level plus the potential value at which they lie. Hence, this energy is perfectly captured by the effective potential of the MLL for states of the first Landau level. 

The first perturbative term beyond the fluctuations of the potential is $\eta\, \Delta V/\beta$. It shows the competition between the fluctuations of the potential and the magnetic length as $\Delta V/\beta \sim \ell_B^2/\ell_{\rm corr}^2$. In the semiclassical regime ($\ell_B\ll \ell_{\rm corr}$) the effective potential is given, up to a small term, by the first Landau level energy plus the potential.

\begin{figure}[h]
	\includegraphics[width=\columnwidth]{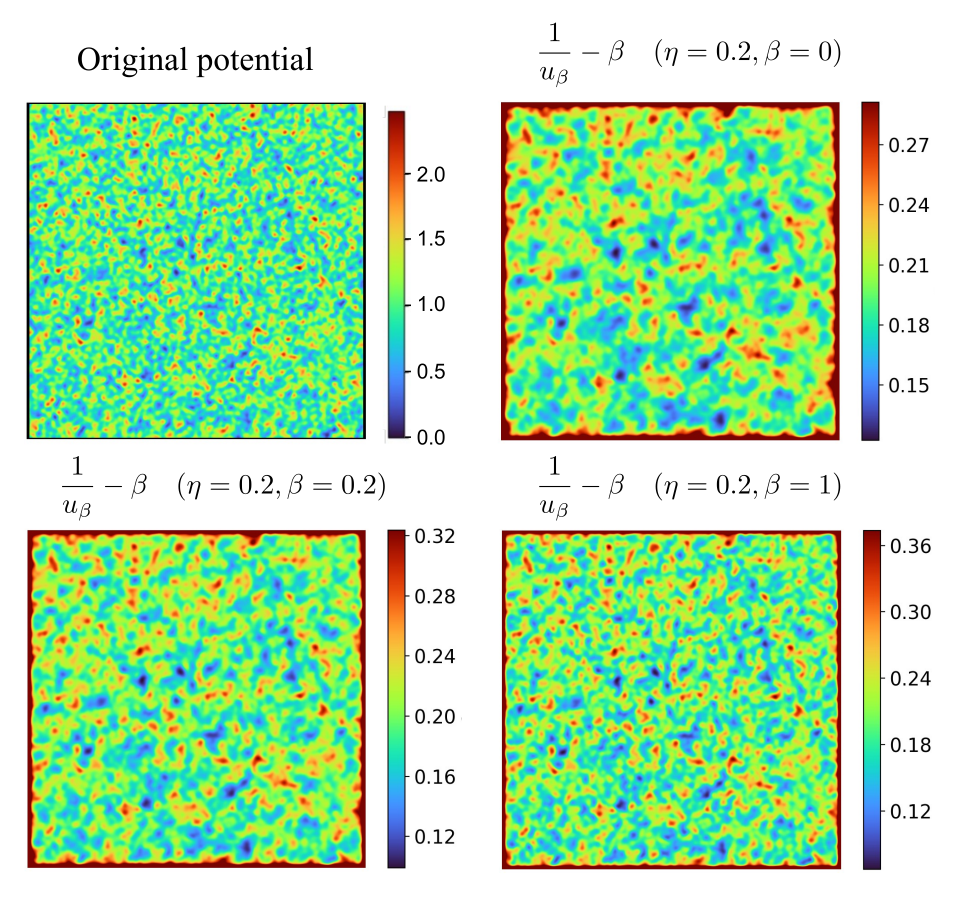}
	\caption{Top left: Original potential~$V$, as defined in Eq.~\eqref{eq:MLL}, mimicking a smoothed compositional disorder. Top right: Effective potential~$1/u$ for $\eta=0.2$, the shift~$\beta$ being equal to zero. Bottom left: Effective potential minus the shift~$\beta$, i.e., $1/u - \beta$, for $\eta=0.2$ and $\beta=0.2$. Bottom right: Effective potential minus the shift~$\beta$ for $\eta=0.2$ and $\beta=1$. As~$\beta$ increases, $1/u-\beta$ converges to $\eta V$.}
	\label{fig:potentials}
\end{figure}

Figure~\ref{fig:potentials} presents a realization of a random potential~$V$ (top left frame) together with the corresponding computed effective potential~$1/u$ obtained from Eq.~\eqref{eq:MLL} for three different sets of values $(\eta,\beta)$: (0.2,0) in the top right frame, (0.2,0.2) in the bottom left frame, and (0.2,1) in the bottom right frame. One sees that in the absence of a magnetic field, the correlation length of the effective potential is larger than the one of the original potential, and that it decreases steadily as $1/u$ converges to $\beta + \eta V$ for large magnetic field, see Eq.~\eqref{eq:semiclassical_expansion}.

%\subsubsection{The general structure of the MLL}

\subsection{Spatial structure of the magnetic states}

\begin{figure}[h]
	\centering
	\includegraphics[width=.49\columnwidth]{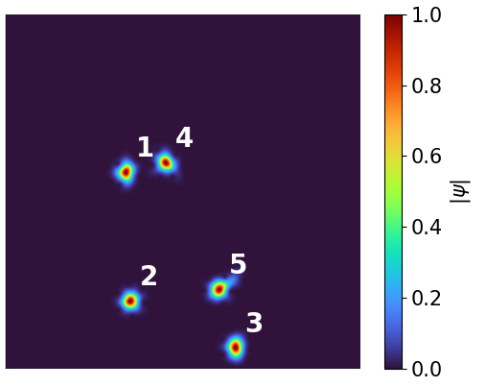}
	\includegraphics[width=.49\columnwidth]{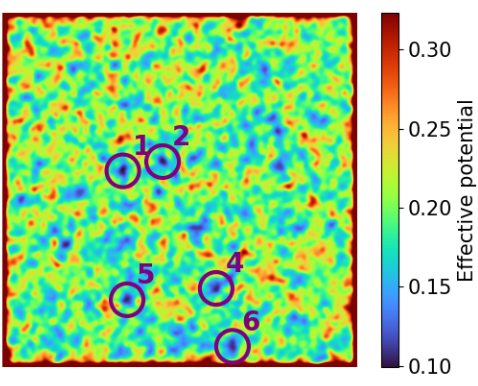}
	\caption{Localization of low-energy states in the wells of the effective potential, for $\eta=0.2$ and $\beta=0.2$. Left: Modulus of the first five eigenfunctions. Right: The effective potential minus $\beta$ with the corresponding wells, indexed by the order of the local minima of the effective potential. The circles indicate the position of the maximal probability density for each eigenstate.}
	\label{fig:firstfive}
\end{figure}

In the non-magnetic LL theory, the low-energy localized eigenstates are located in the basins of the effective potential~\cite{Arnold_2016_PhysRevLett}. Obviously, this is true for the MLL at $\beta=0$ since in this case, the MLL also coincides with the~LL. What is interesting is that this remains true as we increase $\beta$ up to the semiclassical regime. In Fig.~\ref{fig:firstfive} we plot the first five eigenstates for a realization of the random potential and for weak disorder $\eta=0.001$  and the corresponding wells in the effective potential indexed by the order of the corresponding local minima. The order of the eigenstates does not exactly follow the depth of the effective wells. However, we show below that there is a correlation between their energy and the depth of the corresponding well of the effective potential.

\begin{figure}[h!]
	\centering
	\includegraphics[width=\columnwidth]{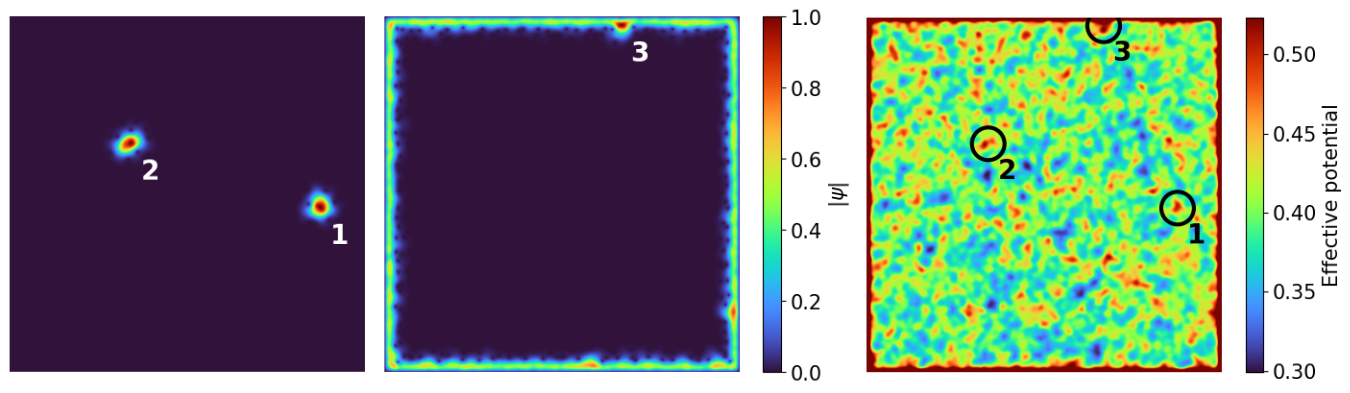}
	\caption{Localization of high-energy states in the hills of the effective potential, for $\eta=0.2$ and $\beta=0.2$. Left: Modulus of two eigenstates localized in the bulk. Middle: Modulus of an eigenstate whose peak is located close to the boundary. We observe a contribution to the probability density near the entire boundary. Right: corresponding effective potential. The circles indicate the position of the maximal probability density for each eigenstate.%The energies are around $\beta+1.5 \eta$.
    }\label{fig:peaks}
\end{figure}

In the upper part of the Landau level, bulk states are well localized around the hills of the effective potential, as displayed in Fig.~\ref{fig:peaks}, where we show two examples of eigenstates located in the bulk of the system. Here we recover a semiclassical picture in which the particle trajectory follows an equipotential line which circles around a minimum of the potential below the critical energy, and around a maximum above the critical energy, except that now the potential has been replaced by the effective potential~$1/u$. However, since the effective potential diverges at the boundary of the domain, this picture loses its validity in its vicinity and the locations of the states cannot be predicted within a characteristic distance~$\ell_e$ to the boundary given by (see the detailed derivation in the Appendix):
\begin{equation}\label{eq:edge_effect}
    \ell_e=\frac{1}{\sqrt{\beta+\eta}}\ln( \frac{2\sqrt{\beta+\eta}}{\eta}) \,.
\end{equation}

\begin{figure*}[ht!]
	\centering
	\includegraphics[width=\textwidth]{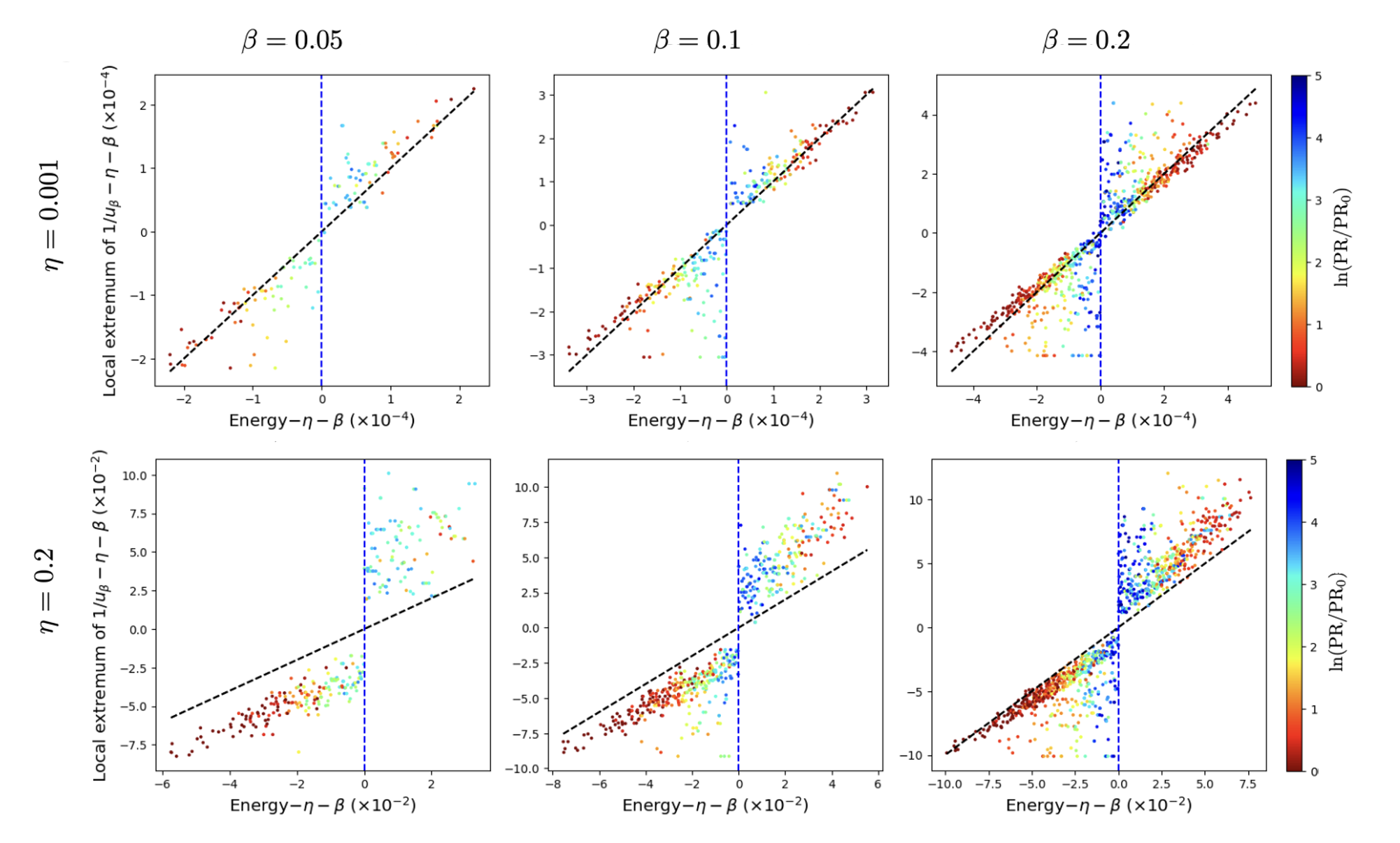}
	\caption{Local extrema of $1/u$ predict the eigenstate energies. For each eigenstate in the first Landau level below (respectively, above) the critical energy $E_c=\eta+\beta$ ---materialized by the vertical dashed line---, the local minimum (respectively, maximum) of the effective potential is obtained by gradient descent (respectively, ascent) from the eigenstate peak position. The obtained local extremum is plotted vs. the state energy. The color of each dot indicates the localized (red) or delocalized (blue) character of the eigenstates, compared to a Gaussian bump of size $\ell_B$. Due to the divergence of the effective potential at the boundary which induces overestimation of the state energy, the states closer to the boundary than a distance~$\ell_e$ are not represented.}
    \label{fig:minmax}
\end{figure*}

\subsection{Spectral structure and energy estimates}

In the semiclassical picture, the eigenstates are located on equipotential contours around the extrema of the potential: around the minima below the critical energy, and around the maxima above. Their energies correspond to the value of the potential on the contour line plus the quantized kinetic energy corresponding to their Landau level. Our MLL approach allows us to generalize this picture and to assess the eigenstate energies from the effective potential. To that end, we systematize the correspondence observed in the previous section and associate a local minimum (respectively, maximum) of the effective potential to each eigenstate below (respectively, above) the critical energy by performing a gradient descent (respectively, ascent) on the effective potential from the point of maximum probability density of the eigenstate. For instance, in Fig.~\ref{fig:firstfive}~(left), the ground state labeled 1 has energy $E_1=0.301$. It is associated to the minimum of the effective potential in Fig.~\ref{fig:firstfive}~(right) designated by a purple circle, also with label~1. The corresponding value of this minimum of the effective potential is $(1/u)_{\rm min}=0.302$.

In Fig.~\ref{fig:minmax} we plot the local minima (resp. maxima) of the effective potential versus the energy for eigenstates below (resp. above) the critical energy and for five realizations of the potential. Note that we have removed from these graphs the eigenstates located too close to the edges of the domain where the divergence of the effective potential leads to a strong deviation of the MLL estimate, the characteristic distance to the edge below which this deviation is significant being $\ell_e$, see Eq.~\eqref{eq:edge_effect} and Appendix.

All eigenstates reported in the graphs have their peak further than $\ell_e$ from the edge of the domain. We observe that when the magnetic intensity is larger than the disorder intensity, i.e., $\beta\gtrsim \eta$ (frames (c)-(f) of Fig.~\ref{fig:minmax}), a large fraction of the points are on the diagonal, which means that the local extrema of the effective potential are good predictors of the eigenstate energies.

\begin{figure*}
	\centering
	\includegraphics[width=0.95\textwidth]{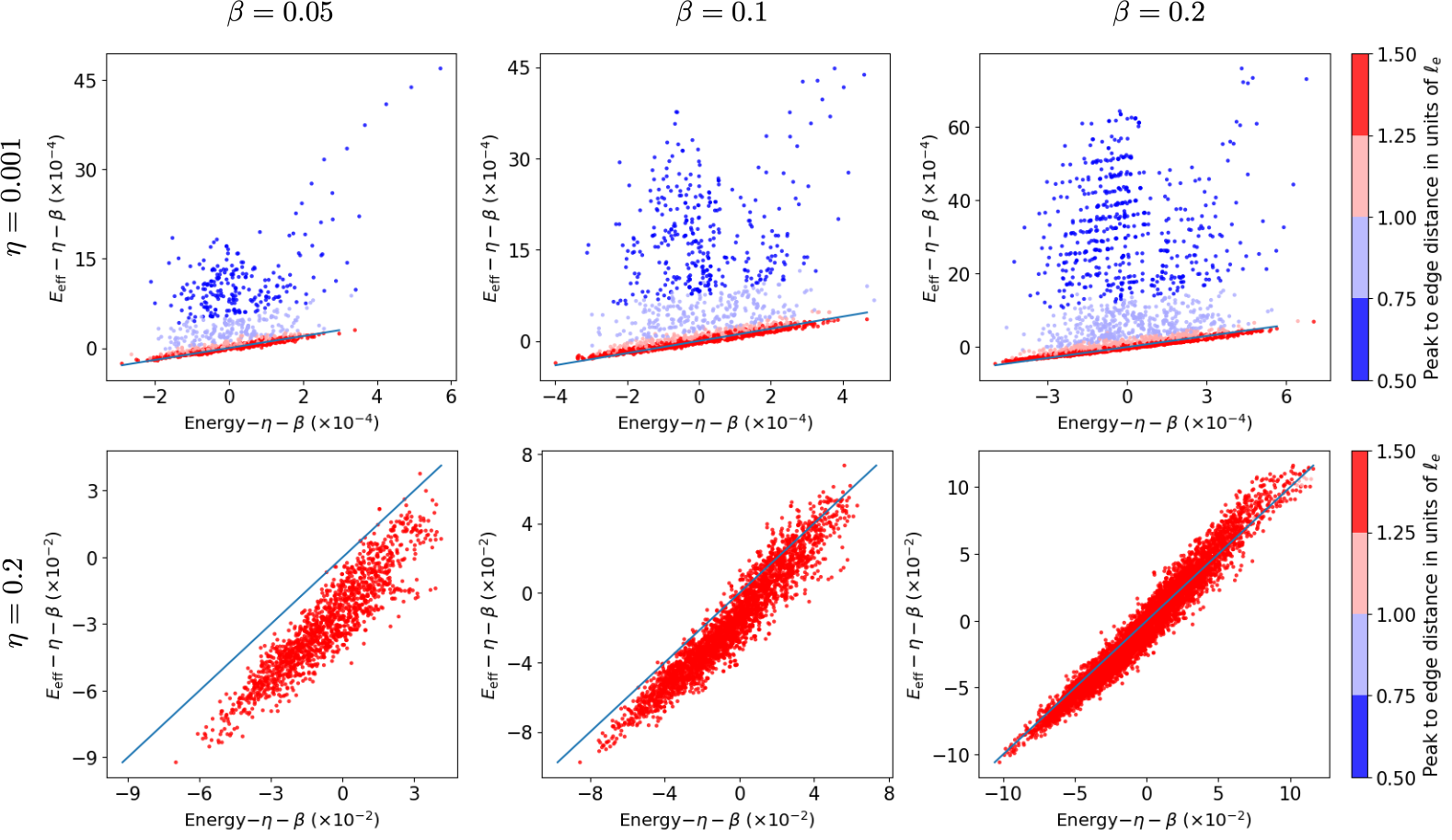}
	\caption{Effective potential value~$E_{\rm eff}$ [see Eq.~\eqref{eq:Eeff}] at the point of maximal probability density of a state vs. the energy~$E$ of the state, for five potential realizations. The colors of the dots indicate the distance of the state to the boundary: red dots correspond to states located further than a distance~$1.25\, \ell_e$ to the boundary [see Eq.~\eqref{eq:edge_effect} and Appendix]. States closer than $0.5\, \ell_e$ to the boundary are not plotted. The blue line inside each frame indicates $E_{\rm eff}=E$. All states in the bulk exhibit a clear linear correlation between $E$ and $E_{\rm eff}$, with $E_{\rm eff} \simeq E$ except for the lower left frame ($\eta=0.2, \beta=0.05$).}
    \label{fig:eff_pot_maxproba}
\end{figure*}

To quantify the spatial extent of an eigenstate~$\psi$, we compute its participation ratio (PR) defined as:
\begin{equation}
	{\rm PR}(\psi )=\frac{ \left(\int \abs{\psi}^2\right)^2}{\int \abs{\psi}^4} \,.
\end{equation}
This quantity is homogeneous to an area and corresponds roughly to the area inside the red line in Figs.\ref{fig:eigenfunction_types}(b) and~\ref{fig:eigenfunction_types}(d). Note that this quantity differs from the localization length because it represents the size of the main region of existence of the wave function, whereas the localization length characterizes the spatial decay of the wave function away from this region. We find that the least extended states have a typical PR of order $\ell_B^2$, which is the one of a Gaussian state~$\psi_0(\vb{r})\propto \mathrm e^{-r^2/4\ell_B^2}$ of the magnetic Schrödinger operator in the absence of disorder. The PR of this Gaussian state is precisely 
\begin{equation}
    {\rm PR}_0 = 4\pi \, \ell_B^2 = \frac{4 \pi}{\beta}\,,
\end{equation}
where the last equality holds in dimensionless units. In Fig.~\ref{fig:minmax}, the ratios~${\rm PR}/{\rm PR}_0$ of all states are color-coded in log scale. We observe a very good correspondence between the MLL prediction (i.e., the value of the corresponding extrema of $1/u$) and the actual energy for all eigenstates having a ratio ${\rm PR}/{\rm PR}_0$ smaller than 2. These more localized states are located exactly on the wells and the hills of the effective potential of the MLL. For each of these states, the local extremum of the effective potential almost coincides with the peak of the eigenstate, and one can directly read the energy of the state by looking at the value of the effective potential.
 
Finally, more extended states in the center of the Landau level do not sit particularly in a single well or on a single hill. Yet, for these states the value of the effective potential at the peak of the eigenstate is still very well correlated to its energy, even if it does not correspond anymore to a maximum or a minimum. In Fig.~\ref{fig:eff_pot_maxproba} we plot the value of the effective potential at the peak of each eigenfunction~$\psi$, called here $E_{\rm eff}(\psi)$, as a function of the actual computed energy of $\psi$, $E(\psi)$, for five disorder realizations:
\begin{equation}\label{eq:Eeff}
    E_{\rm eff}(\psi) = \frac{1}{u\left(\arg \displaystyle \max_{\vb{r}}(\abs{\psi(\vb{r})}\right)} \,.
\end{equation}
Eigenstates with peaks in the bulk are represented by red points, and those closer to the edge in blue, the transition from red to blue occurring at a distance of the order of~$\ell_e$.  All points representing eigenstates in the bulk of the system align well on the identity line (solid blue line) when $\eta\lesssim \beta$, showing a striking correspondence between $E_{\rm eff}(\psi)$ and $E(\psi)$. Of course, in these cases the MLL estimate cannot be called a prediction of the energy in itself as it requires the \textit{a priori} knowledge of the location of the peak of the eigenfunction. Yet, this shows the aptitude of the MLL at capturing the spatial and spectral structure of the electronic states.

\section{Discussion}

Our approach captures in one single scalar function, namely the MLL~$u$ or more precisely its reciprocal, the effective potential~$1/u$, a wide range of spatial and spectral properties of the eigenstates of the magnetic Schr\"odinger operator. This effective potential generalizes and extends the semiclassical approach even in a range of parameters where one would be \textit{a priori} far from the semiclassical regime. The magnetic eigenfunctions are located around the minima of $1/u$ below the critical energy (corresponding to classical orbits in the basins of $1/u$), while they are located around the maxima of $1/u$ above the critical energy. Below the critical energy corresponding to the center of the Landau level, the localization length increases with energy to eventually diverge at the critical energy. Above this critical energy, the localization length decreases while some wave functions have a significant part of their probability density located near the edges of the domain. In both cases, the locations of the wells and hills of the effective potential, together with their sizes, depths, and heights, are good predictors of the locations, sizes and energies of the corresponding localized eigenstates.

However, unlike the non magnetic case the energy of a magnetic state cannot be reformulated here exactly through the MLL as in Eq.~\eqref{eq:conjugate_energy} (see also Ref.~\citep[Eq.~(6)]{Arnold_2016_PhysRevLett}) and only a control inequality remains, see~Eq.~\eqref{eq:energy_inequality}. Therefore, one cannot define an Agmon distance similar to the one expressed for the non magnetic case, see Eq.~\eqref{eq:Agmon_non_mag}, which would finely control the long-range decay of the localized states. Despite this limitation, the prediction of the localization regions and the energy estimates provided by the MLL appear reliable in a large fraction of the $(\beta, \eta)$ parameter space where the semiclassical approach cannot be used.

Since the MLL is the solution to a single Dirichlet problem, the information it provides about the magnetic eigenstates is not statistical but specific to each realization of disorder, irrespective of the disorder type, including long-range or correlated~\cite{Cain_2001_PhysRevB}. Its predictions could therefore be directly compared with recent scanning tunneling microscopy observations~\cite{Bindel_2017_PhysRevLett}. The MLL also provides all the necessary ingredients to compute the conductivity of individual system in a Chalker-Coddington model~\cite{Chalker_1988_JPhysC, Shaw_2025_PhysicaE}. In this framework, the random conductivity network derived from the MLL would be the one formed by the wells and the hills of the effective potential $1/u$, while the saddle points of the same effective potential would determine the quantum tunneling between magnetic states. Analyzing the statistical properties of~$1/u$ would make it possible to infer the system’s transport properties and the associated critical exponents.

\section{Conclusion}

In this work, we have introduced and applied the MLL, a version of the localization landscape which includes magnetic effects, based on the mathematical scheme proposed by Poggi~\cite{Poggi_2024_AdvMath}. Using numerical simulations of eigenstates of a QHE system in the continuous setting with a bounded correlated disorder, we have shown that the topography of the MLL’s effective potential captures key features of the localization properties of the magnetic states. In particular, we were able to recover a semiclassical picture in which these eigenstates are localized around the wells of the effective potential associated to the MLL below the critical energy, while they are located around its hills above. Moreover, the eigenstate energies are linearly correlated to local minima and maxima of the MLL, respectively below and above the critical energy. As a single deterministic function that encodes disorder-specific behavior, the MLL offers a powerful framework for analyzing localization and transport in two-dimensional disordered magnetic systems, and for future investigations of conductivity and critical phenomena in quantum Hall settings.

\section*{Acknowledgments}

We are grateful to B. Poggi, J.-P. Banon, C. Texier, and J.-N. Fuchs for many fruitful discussions. This work was supported by the project Localization of Waves of the Simons Foundation (Grants No.~601944 and No.~1027116, MF).

%\nocite{*}

%\bibliographystyle{apsrev4-2}
\bibliography{magnetic_landscape}

\newpage

\section*{Appendix}

\subsection*{Influence of the boundary on the MLL}

Suppose that the domain is the right half-plane ${\lbrace x>0 \rbrace}$, with Dirichlet boundary condition at $x=0$. If the non dimensional potential~$V$ is constant and equal to 1, then the MLL equation~\eqref{eq:MLL} reads:
\begin{equation}
	-\Delta \bar{u} + \left(\eta+\beta\right) \bar{u} = 1 \,,
\end{equation}
and its solution is:
\begin{equation}\label{eq:ubar_boundary}
	\bar{u}(x)= \frac{1}{\eta + \beta} \left[1-\exp(-\sqrt{\eta + \beta}\,x) \right] \,.
\end{equation}
We see that the effective potential of this MLL vanishes at the boundary and reaches a plateau value $(\eta + \beta)$ in the bulk at a typical distance $1/\sqrt{\eta + \beta}$. We need to compare this influence of the boundary with the typical size of the fluctuations of the effective potential induced by the disorder inside the bulk.

To that end, using a Lippman-Schwinger approach we first show that, if the original potential oscillates locally around a value~$V_0$, $V = V_0 + \delta V$, then the effective potential $1/u$ can be seen as a local averaging of $V$ at the scale $1/\sqrt{V_0}$. In that case, the landscape equation $-\Delta u + Vu = 1$ becomes:
\begin{equation}\label{eq:lansdcape_V0}
    -\Delta u + \frac{u}{\ell_0^2} = 1 - \delta V \, u \,,
\end{equation}
where $\ell_0 = 1/\sqrt{V_0}$ is a characteristic length. The Green function $G(\vb{r},\vb{r}')$ of the differential operator $-\Delta + 1/\ell_0^2$ on the left-hand side takes the following form in $\mathbb{R}^2$:
\begin{equation}
    G(\vb{r},\vb{r}') = \frac{1}{2\pi} \, K_0\left(\frac{\abs{\vb{r}-\vb{r}'}}{\ell_0}\right) \,,
\end{equation}
where $K_0$ is the modified Bessel function of the second kind. We now use this expression to invert Eq.~\eqref{eq:lansdcape_V0}:
\begin{align}
    u(\vb{r}) &= \iint \frac{1}{2\pi} \, K_0\left(\frac{\abs{\vb{r}-\vb{r}'}}{\ell_0}\right) \left[ 1 - \delta V(\vb{r}') \, u(\vb{r}') \right] \,{\rm d}\vb{r}' \nonumber\\
    &= \frac{1}{V_0} - \frac{1}{2\pi} \iint K_0\left(\frac{\abs{\vb{r}-\vb{r}'}}{\ell_0}\right) \,\delta V(\vb{r}') \, u(\vb{r}') \,{\rm d}\vb{r}'\,.
\end{align}
This expression is strictly valid in the infinite domain, but we can assume that far from the boundary, the Green's function even in a finite domain takes a similar form. Taking the reciprocal of the above equation to the first order in $\delta V$ and accounting for the fact the integral of $G$ is $\ell_0^2 = 1/V_0$ leads to:
\begin{align}
    \frac{1}{u(\vb{r})} &\approx V_0 \left[ 1 + \frac{\iint K_0\left(\frac{\abs{\vb{r}-\vb{r}'}}{\ell_0}\right) \,\delta V(\vb{r}') \, \frac{1}{V_0} \,{\rm d}\vb{r}' }{\iint K_0\left(\frac{\abs{\vb{r}-\vb{r}'}}{\ell_0}\right) \,{\rm d}\vb{r}'} \right] \nonumber\\
    &= \frac{\iint K_0\left(\frac{\abs{\vb{r}-\vb{r}'}}{\ell_0}\right) \, V(\vb{r}') \,{\rm d}\vb{r}' }{\iint K_0\left(\frac{\abs{\vb{r}-\vb{r}'}}{\ell_0}\right) \,{\rm d}\vb{r}'} \,.
\end{align}
The above expression shows that $1/u$ corresponds to a local averaging of $V$ at the scale $\ell_0$. Since that more than 70\% of the mass of the denominator is concentrated in a ball of radius $2\ell_0$ around $\vb{r}$, we deduce that far from the boundary the effective potential of the MLL---whose original potential is $\beta + \eta V$---can be seen as a local averaging of this potential on a ball of radius $2/\sqrt{\beta + \eta V}$:
\begin{equation}\label{eq:eff_pot_average}
  \frac{1}{u(\vb{r})} \sim \ev{\beta+\eta V}_{B\left(\vb{r},2/\sqrt{\beta+\eta V}\right)} \,. 
\end{equation}
In our dimensionless MLL equation, the potential $V$ is of order~1 and has a correlation length also of order~1. Therefore, one can remove $V$ in the expression of the ball radius and simply write:
\begin{equation}
    \frac{1}{u(\vb{r})} \sim \beta + \eta \ev{V}_{B\left(\vb{r},2/\sqrt{\beta+\eta}\right)} \,. 
\end{equation}
To assess the amplitude of fluctuations of $\ev{V}$, we use a central limit theorem argument. The correlation length of $V$ being of order 1, the average of $V$ inside a ball of radius~$R$ can be seen as the sum of $N\sim R^2$ independent random variables. The fluctuations of $V$ are therefore of order $1/\sqrt{N} = 1/R$, hence:
\begin{equation}
    \ev{V} = 1 + \frac{\sqrt{\beta+\eta}}{2} \, X(\vb{r}) \,,
\end{equation}
where $X$ is a random function of order 1. Reinserting this expression into~Eq.~\eqref{eq:eff_pot_average} gives:
\begin{equation}
    \frac{1}{u(\vb{r})} = \beta + \eta + \eta \frac{\sqrt{\beta+\eta}}{2} \, X(\vb{r}) \,.
\end{equation}

Using the above equation and Eq.~\eqref{eq:ubar_boundary}, we are now able to compare the modification of $1/u$ due to the presence of a boundary at distance~$x$ and the typical fluctuations of $1/u$ inside the bulk. They are of the same order when:
\begin{equation}
    (\eta + \beta) \left[1 + \exp(-\sqrt{\eta + \beta} \, x)\right] \simeq \beta + \eta + \eta \frac{\sqrt{\beta+\eta}}{2} \, X \,,
\end{equation}
in other words
\begin{equation}
    \exp(-\sqrt{\eta + \beta} \, x) \simeq \frac{\eta}{2\sqrt{\beta+\eta}} \, X \sim \frac{\eta}{2\sqrt{\beta+\eta}} \,. \vspace{3mm}
\end{equation}

This equality occurs at a typical distance $x=\ell_e$ such that:
\begin{equation}\label{eq:elle_appendix}
    \ell_e = \frac{1}{\sqrt{\beta+\eta}} \ln(\frac{2\sqrt{\beta+\eta}}{\eta}) \,.
\end{equation}

\end{document}